\documentclass[aip,reprint,twocolumn,superscriptaddress,groupedaddress]{revtex4-1}  
\usepackage{amsmath}
\usepackage{amssymb}
\usepackage{bbold}
\usepackage{etoolbox}
\usepackage{breqn}
\usepackage{graphicx}
\usepackage{subfigure}
\usepackage{dcolumn}
\usepackage{bm}
\usepackage{float}
\usepackage{hyperref}

\makeatletter
\patchcmd\eq@setnumber{\stepcounter}{\refstepcounter}{}{%
  \errmessage{Patching \noexpand\eq@setnumber failed}%
}
\makeatother
\makeatletter
\let\cat@comma@active\@empty
\makeatother
\let \oldnabla \nabla
\renewcommand{\nabla}{\bm{\oldnabla}}
\renewcommand{\vec}{\mathbf}
\newcommand{\valpha}{\bm \alpha}
\newcommand{\vrho}{\bm \rho}

\newcommand{\halpha}{\hat{\bm\alpha}}

\newcommand{\vom}{\bm \omega}
\newcommand{\haom}{\hat{\bm\omega}}
\newcommand{\om}{\omega}

\newcommand{\vx}{\bm{\sigma}}
\newcommand{\vz}{\vec{z}}
\newcommand{\vv}{\vec{v}}

\newcommand{\scS}{\mathcal{S}}

\newcommand{\uu}[1]{\mathbf{#1}}

\begin{document}

\title{Complexity Reduction Ansatz for Systems of Interacting Orientable Agents: Beyond The Kuramoto Model}
\author{Sarthak Chandra}
\email{sarthakc@umd.edu}
\affiliation{Department of Physics, University of Maryland, College Park, MD 20740, U.S.A.}
\affiliation{Institute for Research in Electronics and Applied Physics, University of Maryland, College Park, MD 20740, U.S.A.}
\author{Michelle Girvan}
\affiliation{Department of Physics, University of Maryland, College Park, MD 20740, U.S.A.}
\affiliation{Institute for Physical Science and Technology, University of Maryland, College Park, MD 20740, U.S.A.}
\author{Edward Ott}
\affiliation{Department of Physics, University of Maryland, College Park, MD 20740, U.S.A.}
\affiliation{Institute for Research in Electronics and Applied Physics, University of Maryland, College Park, MD 20740, U.S.A.}
\affiliation{Department of Electrical and Computer Engineering, University of Maryland, College Park, MD 20742, U.S.A.}
\begin{abstract}
Previous results have shown that a large class of complex systems consisting of many interacting heterogeneous phase oscillators exhibit an attracting invariant manifold. This result has enabled reduced analytic system descriptions from which all the long term dynamics of these systems can be calculated. Although very useful, these previous results are limited by the restriction that the individual interacting system components have one-dimensional dynamics, with states described by a single, scalar, angle-like variable (e.g., the Kuramoto model). In this paper we consider a generalization to an appropriate class of coupled agents with higher-dimensional dynamics. For this generalized class of model systems we demonstrate that the dynamics again contain an invariant manifold, hence enabling previously inaccessible analysis and improved numerical study, allowing a similar simplified description of these systems. We also discuss examples illustrating the potential utility of our results for a wide range of interesting situations.
\end{abstract}
\maketitle
\begin{quotation}
The dynamics of systems with many coupled dynamical agents is a subject of increasing importance with a very broad range of applications. Much of the progress in this field has flowed from the discovery of solvable paradigmatic `toy' systems. In many such systems, the agents are assumed to have one-dimensional dynamics, and a certain class of such systems has been shown to be in some sense `solvable' via a novel analytic technique. 
In our work we consider a more general class of dynamical systems of coupled agents that may have arbitrary dimension. This is a significantly broader class of systems that contains, but is not limited to the previously described class of systems with one-dimensional agents. We then demonstrate that in this broader class of dynamical systems we can construct an analogous technique that can be used to solve such systems. Our method provides analytic techniques to allow previously inaccessible mathematical analyses of these systems. We give significant examples applying our method to large systems of interacting higher-dimensional agents, with a particular focus on the Kuramoto model generalized to higher dimensions.
\end{quotation}

\section{Introduction}

Models of systems of many coupled dynamical agents are useful tools for studying a very wide variety of phenomena\cite{Pikovsky2003}. Examples include flashing fireflies\cite{Ermentrout1991, Buck1968}, circadian rhythms of mammals\cite{Antonsen2008, Childs2008}, oscillating neutrinos\cite{Pantaleone1998}, arrays of Josephson junctions\cite{Marvel2009}, oscillation of footbridges\cite{Abdulrehem2009}, biochemical oscillators\cite{Yamaguchi2003, Kiss2002}, power-grids\cite{Carreras2004, Motter2013}, collections of neurons\cite{Luke2013, Pazo2014, Montbrio2015,Chandra2017}, flocking dynamics\cite{Ha2010,Moshtagh2007,Zhu2013a,Wang2005} and others. In many cases the states of the individual agents can be described by a single angle-like variable, $\theta$. This class of model systems includes situations for which the dynamical agents are oscillators\cite{Pikovsky2003}, neurons\cite{Luke2013, Pazo2014, Montbrio2015,Chandra2017} or robots moving on a two-dimensional plane\cite{Moshtagh2007}, among others. Many such models, possibly involving network-based interactions\cite{Restrepo2014,Abrams2008,Chandra2017}, such as the Kuramoto model\cite{Kuramoto1975}, the Kuramoto-Sakaguchi model\cite{Sakaguchi1986,Abrams2008}, and models of theta neurons\cite{Luke2013} among others\cite{Ott2009}, reduce to the form
\begin{equation}\label{eq:Htcomplex} 
\dot \theta_i = \omega(\eta_i,\{\theta\},t) + \frac{1}{2\iota} [H(\eta_i,\{\theta\},t) e^{-\iota \theta_i} - H^*(\eta_i,\{\theta\},t) e^{\iota \theta_i}], 
\end{equation} 
where $\theta_i$ represents the state of the $i^{\text{th}}$ agent, $\eta_i$ is a (possibly vector) constant parameter that is associated with the $i^{\text{th}}$ agent, $\omega(\eta_i,\{\theta\},t)$ is its ``natural frequency'', $N$ is the total number of agents, and $H(\eta_i,\{\theta\},t)$ is a common field that acts on each agent, dependent on the agent's parameter $\eta_i$, and $\{\theta\}$ indicates a dependence on the set of states $\{\theta_1,\hdots,\theta_N\}$ in the form of an average over $i$ of a function of the angle $\theta_i$.

For example, the well-studied Kuramoto model\cite{Kuramoto1975,Kuramoto1984} can be expressed in the form of Eq. (\ref{eq:Htcomplex}) by choosing $H(\eta_i,\{\theta\},t) = N^{-1}\sum_j \exp(\iota \theta_j)$, independent of $\eta_i$, and choosing $\om(\eta_i,\{\theta\},t)$ to independent of $\{\theta\}$ and $t$, allowing $\om(\eta_i,\{\theta\},t)$ to be replaced by $\om_i$.
Reference \onlinecite{Ott2008} introduced an ansatz to analytically achieve substantial reductions in the complexity of problems of the type exemplified by Eq. (\ref{eq:Htcomplex}) in the limit of a large number of agents ($N\to\infty$). Subsequently, this reduction has been applied in studies of a wide variety of systems (e.g., Refs.\onlinecite{Childs2008,Marvel2009,Abdulrehem2009,Luke2013, Pazo2014, Montbrio2015,Chandra2017,Ottino2018,Roulet2016}).

Several flocking models employ the Kuramoto model (e.g., Refs. \onlinecite{Ha2010,Moshtagh2007,Zhu2013a}) to describe orientational alignment of the velocities of individuals in a flock. Since the standard Kuramoto model (in common with other models conforming to the general form of Eq. (\ref{eq:Htcomplex})) describes the dynamics of scalar angles, these models are restricted to describing flock dynamics in a two-dimensional plane. Other work has shown that the Kuramoto model can be generalized to flocks moving in three and higher-dimensional space\cite{Olfati-Saber2006,Zhu2013,Zhu2014,Lohe2009,Lohe2018,Tanaka2014}. 
In this case each agent's state is assumed to be specified by a unit vector $\vx_i(t)$ in the  $D$-dimensional space. Alternately we may think of $\vx_i$ as specifying a point on the unit sphere in $D$-dimensional space. Reference \onlinecite{Olfati-Saber2006} notes that the vector $\vx_i$ can be thought of as representing the opinion of an individual in a group, or the orientation of the velocity of a member of a flock. (For the case of flocking of birds, fish or flying drones, the generalization to $D=3$ is of most interest.) For $D=2$, the unit vector $\vx_i$ is determined by its scalar orientation angle $\theta_i$ specifying a point on the unit circle, thus recovering the previous model, Eq. (\ref{eq:Htcomplex}) (see Sec. \ref{sec:model}). References \onlinecite{Markdahl2016,Li2014,Jadbabaie2004,Sarlette2009} have also studied the Kuramoto model and its generalizations to higher dimensions in the contexts of continuous-time consensus protocols, multi-agent rendezvous, distributed control, and coalition formation. In this paper we present a new technique that enables analytic treatment of the dynamics of a large class of systems with higher-dimensional agents, including the aforementioned systems. In particular, in this paper we focus on the continuum limit of infinitely many higher-dimensional agents, allowing us to use ideas similar to those developed previously in the context of Eq. (\ref{eq:Htcomplex}).

The remainder of the paper is organized as follows: In Sec. \ref{sec:model} we construct a generalization of Eq. (\ref{eq:Htcomplex}) to arbitrary dimensions and describe the infinite system size limit in such systems. Then, in Sec. \ref{sec:ansatz} we extend the ansatz of Ref. \onlinecite{Ott2008}, resulting in a simplified analytic description of this generalized class of systems. In Sec. \ref{sec:examples} we demonstrate the utility of our results to example systems, with particular focus on the Kuramoto model generalized to higher dimensions.
Finally, in Sec. \ref{sec:conclusions} we conclude with a discussion and summary of our results.

\section{Generalizing Kuramoto-like Agents to Higher Dimensions}\label{sec:model}


In two recent papers\cite{Chandra2019,ChandraIMR}, we constructed a generalization of the Kuramoto model to $D$ dimensions. Here we consider an even more general setup, where we consider a generalization to Eq. (\ref{eq:Htcomplex}) to a system in $D$ dimensions,
\begin{equation}\label{eq:model}
\dot{\vx}_i = [\vrho(\eta_i,\{\vx\},t) - (\vx_i \cdot \vrho(\eta_i,\{\vx\},t))\vx_i] + \uu{W}(\eta_i,\{\vx\},t)\vx_i,
\end{equation}
where for each $i$, $\vx_i(t)$ is a real $D$-dimensional unit vector, $|\vx_i(0)|=1$, $\vrho(\eta_i,\{\vx\},t)$ is an arbitrary real $D$-dimensional vector, which can be thought of as a common field that affects each agent in an $\eta_i$ dependent fashion, $\uu{W}(\eta_i,\{\vx\},t)$ is a real $D\times D$ antisymmetric matrix, $\eta_i$ is a (possibly vector) constant parameter associated with each agent, and, as earlier, $\{\vx\}$  indicates a dependence on the set of all states $\{\vx_1,\hdots,\vx_N\}$ in the form of the average over $i$ of a function of the unit vectors $\vx_i$ (we further quantify this dependence on $\{\vx\}$ later). For example, in the context of flocking agents in $D$ dimensions, $\vx_i$ represents the orientation of the $i^{\text{th}}$ agent, $\vrho(\eta_i,\{\vx\},t)$ represents a `goal' orientation to which the $i^{\text{th}}$ agent attempts to align itself, and $\uu{W}(\eta_i,\{\vx\},t)$ represents a fixed bias, or a systematic error to the agent dynamics causing the agent to head in a direction that deviates from the direction of $\vrho$\cite{Chandra2019}.
Note from the form of Eq. (\ref{eq:model}) that the dot product of the right-hand side of Eq. (\ref{eq:model}) with $\vx_i$ is identically zero, so that $d|\vx|/dt=0$, as required by our identification of $\vx$ as a unit vector. 
Thus the dynamics of each $\vx_i$ is restricted to the $(D-1)$-dimensional surface, $\scS$, of the unit sphere, $|\vx|=1$. For $D=2$, choosing $\vx_i = (\cos\theta_i,\quad \sin\theta_i)^T$, $\vrho(\eta_i,\{\vx\},t) = \left(\text{Re}[H(\eta_i,\{\theta\},t)],\quad \text{Im}[H(\eta_i,\{\theta\},t)]\right)^T$ and 
\begin{equation*}
\uu{W}(\eta_i,\{\vx\},t) = \begin{pmatrix} 0 & \omega(\eta_i,\{\theta\},t) \\ -\omega(\eta_i,\{\theta\},t) & 0\end{pmatrix},
\end{equation*}
reduces Eq. (\ref{eq:model}) to Eq. (\ref{eq:Htcomplex}), thus justifying Eq. (\ref{eq:model}) as a $D$-dimensional generalization of Eq. (\ref{eq:Htcomplex}).

We now consider the limit of a large number of agents, and denote by $F(\vx,\eta,t)$ the distribution of agents on $\scS$, such that $F(\vx,\eta,t)d^{D-1}\sigma d\eta$ is the fraction of agents that lie in the $(D-1)$-dimensional differential element $d^{D-1}\sigma$ on the surface $\scS$ centered at $\vx$ at time $t$, and have an associated parameter $\eta$ within the differential element $d\eta$ centered at $\eta$. Since the associated parameter $\eta$ for each agent is time independent, we define
\begin{equation*}
g(\eta) = \int_{\scS} F(\vx,\eta,t) d^{D-1}\sigma,
\end{equation*}
and 
\begin{equation*}
f(\vx,\eta,t)=F(\vx,\eta,t)/g(\eta).
\end{equation*}

Noting that Eq. (\ref{eq:model}) specifies the vector field of the flow controlling the dynamics of the distribution $f$, we write a continuity equation for $f$,
\begin{equation}\label{eq:Ddimcontinuity}
\partial f(\vx,\eta,t)/\partial t + \nabla_{\scS} \cdot [f(\vx,\eta,t) \vv(\vx,\eta,t)] = 0,
\end{equation}
where the velocity field $\vv(\vx,\eta,t)$ is given by $\vv(\vx,\eta,t) = (\vrho(\eta,t) - (\vx\cdot\vrho(\eta,t))\vx) + \uu{W}(\eta,t)\vx$, and $\nabla_{\scS}\cdot\vec A$ represents the divergence of a vector field $\vec A$, along the surface $\scS$. This can be done if the dependence of $\vrho$ and $\uu{W}$ on $\{\vx\}$ can be specified as a functional of $F(\vx,\eta,t)$ that is not explicitly dependent on $\vx$. (A simple example of such a dependence on $\{\vx\}$ would be the average value of $p(\vx_i)$ for some given function $p$	, which can be written as $\int \int p(\vx) F(\vx,\eta,t) d\vx d\eta$.) Following Appendix B of Ref.\onlinecite{Chandra2019}, Eq. (\ref{eq:Ddimcontinuity}) can be rewritten as 
\begin{dmath}\label{eq:continuity}
{{\partial f/\partial t + [\nabla_{\scS} f(\vx,\eta,t) - (D-1)f(\vx,\eta,t)\vx]\cdot \vrho(\eta,t) }} \\+ (\uu{W}(\eta,t) \vx) \cdot \nabla_{\scS} f(\vx,\eta,t) = 0,
\end{dmath}
where $\nabla_{\scS} \Phi$ is the gradient of a scalar field $\Phi$ projected on the surface $\scS$.

\section{Analytic Solution in the Limit of Large Systems}\label{sec:ansatz}

For $D=2$, Refs. \onlinecite{Ott2008, Ott2009} demonstrated that the ansatz that $f(\theta,t)$ is in the form
\begin{equation}\label{eq:ansatzpoisson}
f(\theta,\eta,t) = \frac{1}{2\pi}\frac{1-|\alpha(\eta,t)|^2}{|e^{\iota \theta} - \alpha(\eta,t)|^2}, 
\end{equation}
where $\alpha(\eta,t)$ is a complex scalar function of $\eta$ and $t$, $|\alpha(\eta,0)|<1$, reduces Eq. (\ref{eq:Htcomplex}) to the following $\theta$-independent form
\begin{equation}\label{eq:complexalphaeqn}
\frac{\partial \alpha}{\partial t} + \iota \eta + \frac{1}{2}\left(H^*(\eta,t) \alpha^2(\eta,t) - H(\eta,t)\right) = 0.
\end{equation}
The form Eq. (\ref{eq:ansatzpoisson}) represents an invariant manifold in the space of possible distributions $f$, that satisfy the continuity equation Eq. (\ref{eq:Ddimcontinuity}) for $D=2$.
Furthermore, previous work\cite{Ott2009,Ott2011} has shown that initial conditions for $f$ are attracted to the invariant manifold Eq. (\ref{eq:ansatzpoisson}) for a large class of possible models of the form Eq. (\ref{eq:Htcomplex}). Thus Eq. (\ref{eq:ansatzpoisson}) can be used to greatly simplify the study of the long-term dynamics of these systems.

Here we present an ansatz demonstrating the existence of a similar invariant manifold for Eq. (\ref{eq:Ddimcontinuity}) in any dimension $D$.
Noting that $e^{\iota\theta}$ can be interpreted as a unit vector in the complex plane and that the complex quantity $\alpha$ can similarly be interpreted as a two-dimensional vector of its real and imaginary parts, based on Eq. (\ref{eq:ansatzpoisson}) we posit the following guess for the form of $f(\vx,\eta,t)$ for arbitrary dimension $D$,
\begin{equation}\label{eq:notnormalizedansatz}
f(\vx,\eta,t) = \frac{N_D(\valpha(\eta,t))}{|\vx-\valpha(\eta,t)|^{\beta_D}},
\end{equation}
where $\valpha$ is a real $D$-dimensional vector such that $|\valpha(\eta,0)|<1$, $\beta_D$ is a yet-to-be-determined constant, and $N_D(\valpha)$ is a scalar normalization chosen to ensure that 
\begin{equation}\label{eq:normalizationintegral}
\int_{\scS} f(\vx,\eta,t)d^{D-1}\sigma = 1.
\end{equation}
Inserting Eq. (\ref{eq:notnormalizedansatz}) into the continuity equation in Eq. (\ref{eq:continuity}), we obtain after some algebra, 

\begin{dmath}\label{eq:expandedcontinuitywithbeta}
(1+|\valpha|^2 - 2\valpha\cdot\vx)\partial_t N_D(\valpha) - \beta_D N_D(\valpha) (\valpha\cdot\partial_t \valpha - \vx\cdot\partial_t\valpha) \\+ N_D(\valpha)\{\beta_D(\valpha\cdot\vrho) + [2(D-1)-\beta_D](\valpha\cdot\vx)(\vrho\cdot\vx) \\- (D-1)(\vrho\cdot\vx)(1+|\valpha|^2) - \beta_D\vx\cdot \uu{W}\valpha\} = 0.
\end{dmath}
For our ansatz Eq. (\ref{eq:notnormalizedansatz}) to apply, the above equation must hold for all $\vx$. Focusing on the term in Eq. (\ref{eq:expandedcontinuitywithbeta}) that is quadratic in $\vx$, i.e., $N_D(\valpha) [2(D-1)-\beta_D](\valpha\cdot\vx)(\vrho\cdot\vx)$, since in general $\valpha$ and $\vrho$ will not be zero for all $t$, we require that
\begin{equation}\label{eq:beta}
\beta_D = 2(D-1).
\end{equation}

With $\beta_D$ in Eq.(\ref{eq:notnormalizedansatz}) determined, we now obtain the normalization constant $N_D(\valpha)$.
To perform the integral in Eq. (\ref{eq:normalizationintegral}), without loss of generality we take the vector $\valpha$ to be along the $\hat z$ axis. For an arbitrary point $\vx$ on $\scS$, we denote the angle between $\vx$ and $\hat z$ by $\theta$. In particular, we note that the distance of the point $\vx$ from the $\hat z$ axis is $\sin\theta$. For a coordinate system on the surface $\scS$, we use $\theta$ as one of the coordinates, denoting position with respect to $\hat z$ on the sphere. From the symmetry of $f$ in Eq.(\ref{eq:notnormalizedansatz}) about the direction $\valpha$, we see that the integrals over these remaining coordinates give the surface area $S_{D-1} \sin^{D-2}\theta$ of the $(D-2)$ dimensional surface of a sphere with radius $\sin\theta$ embedded in $(D-1)$ dimensions, where $S_{D-1} = (2\pi)^{(D-1)/2}/\Gamma((D-1)/2)$ is the area of the sphere of unit radius in $D-1$ dimensional space.  Thus Eq. (\ref{eq:normalizationintegral}) becomes
\begin{equation}
1 = S_{D-1} \int_0^{\pi} \frac{N_D(\valpha) \sin^{D-2}\theta d\theta}{(1+|\valpha|^2 - 2|\valpha|\cos\theta)^{D-1}},
\end{equation}
which can be evaluated to give
\begin{equation}
1 = K_D^{-1} \frac{N_D(\valpha)}{(1-|\valpha|^2)^{D-1}},
\end{equation}
where $K_D$ is a constant dependent only on $D$. This results in
\begin{equation}\label{eq:NDalpha}
N_D(\valpha) = K_D (1-|\valpha|^2)^{D-1},
\end{equation}
giving the form of the ansatz for arbitrary dimensions as
\begin{equation}\label{eq:finalansatz}
f(\vx,\eta,t) = K_D \frac{(1-|\valpha(\eta,t)|^2)^{D-1}}{|\vx-\valpha(\eta,t)|^{2(D-1)}},
\end{equation}
which, for $D=2$, agrees with Eq. (\ref{eq:ansatzpoisson}).

To determine whether the ansatz Eq. (\ref{eq:finalansatz}), is consistent with Eq. (\ref{eq:expandedcontinuitywithbeta}) we insert it into Eq. (\ref{eq:expandedcontinuitywithbeta}). We find that the ansatz with $\beta_D$ given by Eq. (\ref{eq:beta}) indeed is a solution of Eq. (\ref{eq:expandedcontinuitywithbeta}) and that Eq. (\ref{eq:expandedcontinuitywithbeta}) reduces to the following equation for $\valpha$ (see Appendix \ref{apx:alphaeqn} for details),
\begin{equation}\label{eq:alphaeqn}
\partial_t \valpha = \frac{1}{2}(1+|\valpha|^2)\vrho - (\vrho\cdot\valpha)\valpha + \uu{W}\valpha.
\end{equation}
The key point is that Eq. (\ref{eq:alphaeqn}) does not involve $\vx$ (and remarkably, also does not involve any dependence on $D$).
Thus, analogously to Eq. (\ref{eq:complexalphaeqn}), we have a $\vx$-independent description of the dynamics of $\valpha$. This is our main result.

We note that for initial conditions with $|\valpha|<1$, $|\valpha|$ will remain less than 1 at all finite times since from Eq. (\ref{eq:alphaeqn}) $\partial_t |\valpha|=0$ at $|\valpha|=1$, thus verifying that $f$ given by Eq. (\ref{eq:finalansatz}) does not diverge for $t<\infty$.

\section{Example Systems}\label{sec:examples}

We now consider a few examples illustrating the utility of the generalized ansatz, Eq. (\ref{eq:finalansatz}), to systems of the form given in Eq. (\ref{eq:model}). We detail the particular example of the Kuramoto model generalized to $D$ dimensions\cite{Chandra2019} as representative of the utility of our main result Eq. (\ref{eq:alphaeqn}), and thereafter briefly mention applications of this result to a variety of other systems.

\subsection{The Kuramoto Model Generalized to Higher Dimensions}\label{sec:genkura}

A generalization of the Kuramoto model with homogenous oscillators to arbitrary dimension was introduced by Olfati-Saber in 2006\cite{Olfati-Saber2006} in the context of flocking dynamics, consensus protocols, and opinion dynamics. This was later generalized to heterogeneous systems by Chandra et al.\cite{Chandra2019}. For generalization to $D$ dimensions, a system order parameter, $\vz$, can be defined as
\begin{equation}\label{eq:ordpar}
\vz(t) = \frac{1}{N}\sum_i \vx_i(t).
\end{equation}
The magnitude of $\vz(t)$ is a measure of the coherence of the set of agents $\{\vx\}$.
The common field $\vrho$ is then defined as the $\eta_i$-independent function,
\begin{equation}\label{eq:kuramotorho}
\vrho(\eta,\{\vx\},t) = K \vz(t) = (K/N)\sum_i \vx_i(t),
\end{equation}
where $K$ is a coupling constant. By interpreting the vector parameters $\eta_i$ in $\uu{W}(\eta_i,\{\vx\},t)$ as the $D(D-1)/2$ independent elements of a $D$-dimensional antisymmetric matrix $\uu{W}_i$, we can replace $g(\eta)d\eta$ in integrals by $G(\uu{W})d\uu{W}$ where $G(\uu{W})$ is a distribution of antisymmetric matrices. In cases such as these where $\uu{W}(\eta_i,\{\vx\},t)$ is independent of $\{\vx\}$ and $t$, we interpret $\uu{W}(\eta_i)=\uu{W}_i$ as the ``natural rotation'' of $\vx_i$.

In the limit of infinite system size, with a distribution of agents given according to Eq. (\ref{eq:finalansatz}), 
\begin{dgroup*}
\begin{dmath*} \vz(t) = \int_{\scS} F(\vx,\uu{W},t)\vx d^{D-1}\sigma d\uu{W}, \end{dmath*}
 \begin{dmath} = \int d\uu{W} G(\uu{W}) \valpha(\uu{W},t)/|\valpha(\uu{W},t)|\\ \times \int_0^{\pi} \frac{K_D (1-|\valpha(\uu{W},t)|^2)^{D-1} \cos\theta  \sin^{D-2}\theta d\theta}{(1+|\valpha(\uu{W},t)|^2 - 2|\valpha(\uu{W},t)|\cos\theta)^{D-1}} \label{eq:kuramotozintegral} \end{dmath}
\end{dgroup*}

For $D=2$ (i.e, the original Kuramoto model) Eq. (\ref{eq:kuramotozintegral}) evaluates to give $\vrho(t)=K\vz(t)=K\int d\om g(\om) \valpha(\om,t)$. Equation (\ref{eq:alphaeqn}) is then equivalent to Eq. (6) from Ref.\onlinecite{Ott2008}. For $D=3$, the integral in Eq. (\ref{eq:kuramotozintegral}) gives
\begin{dmath}\label{eq:D3rhointegral}
\vrho = K\int d\uu{W} G(\uu{W}) \valpha(\uu{W},t)/|\valpha(\uu{W},t)|\\ \times \left. \left[2|\valpha|(1+|\valpha|^2) + (1-|\valpha|^2)^2 \log\left(\frac{1-|\valpha|}{1+|\valpha|}\right)\right] \middle/ 4|\valpha|^2 \right. .
\end{dmath}
This now allows us to use Eq. (\ref{eq:alphaeqn}) with some given $G(\uu{W})$ to numerically integrate for the dynamics of $\valpha$, and the dynamics of the order parameter $\vrho$.

\begin{figure}
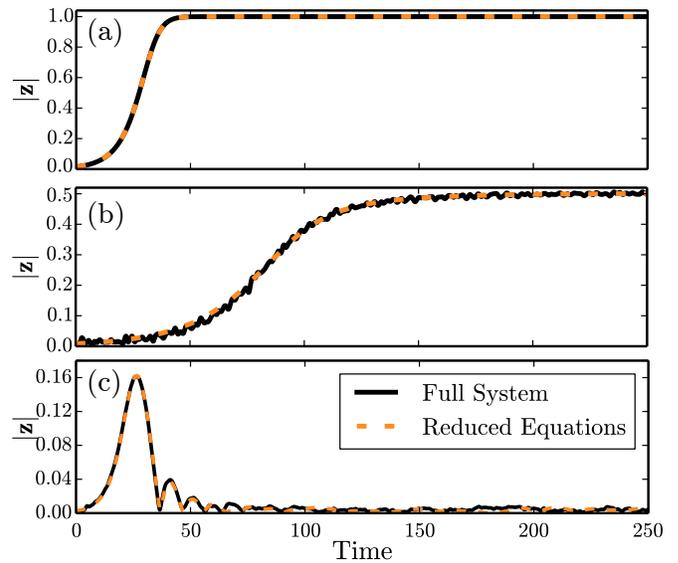

\includegraphics[width=\columnwidth]{{{Ansatz_full_system_match_w_IMR}}}
\caption{(a),(b): Comparison between the dynamics of the magnitude of the order parameter, $|\vz|$, as a function of time via full system modeling of the generalized Kuramoto model with $D=3$ (Eq. (\ref{eq:model}) for $\vrho$ given by Eq. (\ref{eq:D3rhointegral})) using $N=5000$ agents shown in black, with the modeling of the reduced differential equation Eq.(\ref{eq:alphaeqn}) plotted as the orange dashed line. $K=2$ for both figures. (a) is the case of homogenous agents, i.e., $G(\uu{W})=\delta(\uu{W}-\uu{W_0})$. (b) is the case of heterogeneous agents, where the distribution $G(\uu{W})$ is nonsingular and chosen as described in the main text. Only $N_{\uu{W}}=500$ Monte-Carlo samples were required to produce the curve for the reduced system of equations, representing the $N\to\infty$ limit of the full system, approximated by the noisy curve generated using $N=5000$ agents for the full system. 
(c) demonstrates similar agreement for the case of heterogeneous agents in $D=4$, where the system is evolved at $K=1.7$ from the uniform incoherent distribution as the initial condition. $N=N_{\uu{W}}=10^6$ was used for numerical integration of the two curves. Note how the reduced equations capture the transient behavior of the Instability-Mediated Resetting phenomenon\cite{ChandraIMR} (discussed in text).
Since the initial finite-size noise is different in the two cases, in order to make the curves for the full system and the reduced equations lie on each other, we shift them in time to align them. See text for further details of initial conditions used. 
}
\label{fig:numericstheoryfit}
\end{figure}

Using this simplification, we can efficiently simulate the dynamics of the full system of agents governed by Eq. (\ref{eq:model}). 
We first focus on the case of homogenous agents, i.e., identical natural rotations for each agent, $G(\uu{W})=\delta(\uu{W}-\uu{W_0})$, where $\delta(\cdot)$ is the Dirac-delta function. We can then change to a rotating basis in which the natural rotation term of each agent is zero, $\uu{W_0}\to 0$. 
This makes the $\uu{W}$-integral in Eq. (\ref{eq:kuramotozintegral}) trivial, allowing a direct representation of $\vrho$ in terms of $\valpha$. Further, $\valpha$ is only dependent on time (rather than $\uu{W}$ and $t$). This represents a very large simplification in the complexity of the dynamics of the system of agents, since Eq. (\ref{eq:alphaeqn}) is now a single $D$-dimensional ordinary differential equation which represents the collective dynamics of the $N\to\infty$, $D$-dimensional system of coupled differential equations in Eq. (\ref{eq:model}). The utility of this result is demonstrated for $D=3$ in Fig. \ref{fig:numericstheoryfit}(a), where we show (plotted in black) the time-series for $|\vrho(t)|$ as generated from a system of $N=5000$ agents (approximating the $N\to\infty$ limit), compared with the time-series generated from the theory derived in Eq. (\ref{eq:alphaeqn}) (orange dashed curve). The initial condition for the full system was chosen such that the agents were uniformly randomly distributed on the sphere. For the theory derived in Eq. (\ref{eq:alphaeqn}), i.e., the reduced equations, the initial value of $\valpha$ was chosen to have magnitude $0.01$ in an arbitrary direction. Note the remarkably close agreement between the black and the orange dashed curve, demonstrating that the dynamics on the reduced manifold of Eq. (\ref{eq:finalansatz}) indeed gives the large-$N$ dynamics of the full system of interacting agents. 

 For the case of heterogeneous agents, $\valpha$ in Eq. (\ref{eq:kuramotozintegral}) depends on $\uu{W}$, and we perform the integral in a Monte-Carlo fashion. We randomly choose $N_{\uu{W}}$ values of $\uu{W}$ from the given distribution $G(\uu{W})$ and simulate the dynamics of the corresponding $\valpha(\uu{W})$s. These randomly chosen $\valpha(\uu{W})$s are then used as the Monte-Carlo samples to evaluate $\vz$ according to Eq. (\ref{eq:kuramotozintegral}), simulating the dynamics of the system in the $N\to\infty$ limit by only simulating the dynamics of $N_{\uu{W}}$ variables. Here we choose an isotropic distribution $G(\uu{W})$ constructed by choosing each upper triangular element from identical independent normal distributions with zero mean and unit variance, and choosing the remaining elements to make $\uu{W}$ antisymmetric.
Results are shown in Fig. \ref{fig:numericstheoryfit}(b) for $D=3$, where $N_{\uu{W}}=500$ Monte-Carlo samples were chosen to evaluate the $|\vrho(t)|$ curve via the theory in Eq. (\ref{eq:alphaeqn}), and are compared with the curve obtained for simulating the dynamics of the full system of equations in the $N\to\infty$ limit, approximated by a simulation of $N=5000$ agents. Note how simulating the dynamics of $N_{\uu{W}}\ll N$ Monte-Carlo samples yields a smooth curve approximating the noisy curve generated by simulating the individual dynamics of $N=5000$ agents. 
Initial conditions for the full system were chosen as a bimodal distribution of $\vx_i$s, independent of the corresponding $\uu{W}_i$, with the two peaks being anti-podal to each other, hence representing a distribution explicitly not on the manifold dictated by Eq. (\ref{eq:finalansatz}). 
The initial condition for the reduced equations, i.e., Eq. (\ref{eq:alphaeqn}) were chosen to be uniform on a sphere of radius $0.01$, corresponding to an approximately uniform distribution of $f(\vx,\eta,t)$ in $\vx$. 
Despite not lying on the invariant manifold described by Eq. (\ref{eq:finalansatz}), we observe that the dynamics of the full system rapidly approach the dynamics as predicted by Eq. (\ref{eq:alphaeqn}) for the $N\to\infty$ limit for dynamics on the invariant manifold. This indicates that for the case of heterogeneous agents the invariant manifold Eq. (\ref{eq:finalansatz}) is attracting, as has been proven for the case of $D=2$\cite{Ott2009}. Full system simulations with initial conditions described by a uniform distribution in $\vx$ (and hence lying on the invariant manifold Eq. (\ref{eq:finalansatz}) for $|\valpha|=0$) yielded a curve that is not discernibly different from the curve presented in Fig. \ref{fig:numericstheoryfit}(b).

\subsection{Applications of Eq. (\ref{eq:alphaeqn}) to previous results on the Generalized Kuramoto model}

\begin{figure}
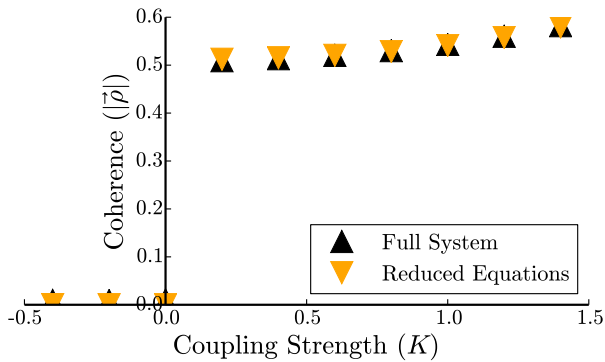

\includegraphics[width=\columnwidth]{{{3D_Phase_transition_ansatz_match}}}
\caption{A simulation of the phase transition to coherence via numerical integration of Eq. (\ref{eq:model}) for $\vrho$ given by Eq. (\ref{eq:D3rhointegral}) representing the full system dynamics with $N=5000$ (shown in the black triangular markers), and via numerical integration of Eq. (\ref{eq:alphaeqn}) representing the dynamics on the invariant manifold with $N_{\uu{W}}=500$ (shown as the orange inverted triangles) for $D=3$. For each value of $K$, the system is evolved until $|\vrho|$ reaches an equilibrium. Note the close agreement between the time asymptotic values of $|\vrho|$ at all values of $K$. The distribution $G(\uu{W})$ was chosen as described earlier for heterogeneous agents.
}
\label{fig:transitionmatch}
\end{figure}

As demonstrated in Fig. \ref{fig:numericstheoryfit}, numerical integration of the dynamics on the invariant manifold, via Eq. (\ref{eq:alphaeqn}) closely reproduces the time evolution of the order parameter of the Kuramoto model generalized to higher dimensions (Eq. (\ref{eq:model}) for $\vrho$ according to Eq. (\ref{eq:D3rhointegral})). As we demonstrate in Fig. \ref{fig:transitionmatch}, this close similarity between a simulation of the full $N$-agent dynamics and the simulation of the reduced equation Eq. (\ref{eq:alphaeqn}) holds at all values of the coupling constant $K$. This allows us to recreate the discontinuous phase transition of the Kuramoto model generalized to 3 dimensions reported in Ref. \onlinecite{Chandra2019} (The continuous phase transition observed through reduced equations of the form Eq. (\ref{eq:alphaeqn}) for the standard Kuramoto model in two dimensions has been demonstrated and discussed in Ref. \onlinecite{Ott2008}). 

Ref. \onlinecite{ChandraIMR} demonstrated that the Kuramoto model generalized to even dimensions $D\geq 4$ exhibits the unusual behavior of Instability-Mediated Resetting: If the coupling strength $K$ is increased abruptly while remaining below the critical coupling strength for the onset to coherence, the Kuramoto model displays a short burst of coherence (see Ref. \onlinecite{ChandraIMR} for further details). This is illustrated by the results shown in Fig. \ref{fig:numericstheoryfit}(c). In Fig. \ref{fig:numericstheoryfit}(c) the initial conditions for the $\vx_i$ in the full system simulation were chosen independently to be uniformly random over the sphere $|\vx|=1$. For the reduced equations, initial conditions for $\valpha$ were chosen similar to the earlier discussion on heterogeneous agents, i.e., uniform on a sphere of radius $0.01$, corresponding to an approximately uniform distribution of $f(\vx,\eta,t)$ in $\vx$. 
Figure \ref{fig:numericstheoryfit}(c) demonstrates that the dynamics of Eq. (\ref{eq:alphaeqn}), representing the dynamics on the invariant manifold described by Eq. (\ref{eq:finalansatz}) (shown in the dashed orange curve), accurately captures these short bursts of coherence, further demonstrating the capability of Eq. (\ref{eq:alphaeqn}) in capturing the transient dynamics of the Kuramoto model generalized to higher dimensions.

To demonstrate the applicability of Eq. (\ref{eq:alphaeqn}) in improving theoretical understanding of such systems, we present an example of a stability analysis of the Kuramoto model generalized to three dimensions via a study of the dynamics on the reduced manifold Eq. (\ref{eq:finalansatz}). In particular, we study the stability of the completely incoherent state for a system of heterogeneous agents, wherein the initial condition of each agent $\vx_i$ is to be distributed independently and uniformly over the sphere $|\vx|=1$. In the $N\to\infty$ limit this is the distribution $F(\vx,\eta,t) = g(\eta)/(4\pi)$ corresponding to setting $|\valpha|=0$ in Eq. (\ref{eq:finalansatz}).
[In the context of the stability analysis presented in Sec. III B of Ref. \onlinecite{Chandra2019}, we are considering the case of $p=0$]. Thus we are interested in the stability analysis of Eq. (\ref{eq:alphaeqn}) about $|\valpha|=0$. Performing a first order expansion of Eq. (\ref{eq:alphaeqn}) for $|\valpha|\ll 1$,
\begin{equation}
\partial \valpha(\uu{W},t)/\partial t = \vrho(t) / 2 + \uu{W} \valpha(\uu{W},t).
\end{equation}
Assuming $\valpha(\uu{W},t) = \valpha(\uu{W})e^{st}$, we obtain
\begin{equation}\label{eq:smallalphasoln}
\valpha(\uu{W}) = \frac{1}{2}(s\mathbb{1} - \uu{W})^{-1} \vrho.
\end{equation}
Note that in the limit of $|\valpha|\ll 1$, Eq. (\ref{eq:D3rhointegral}) can be written as 
\begin{equation}
\vrho = \left(\frac{4K}{3}\right) \int d\uu{W} G(\uu{W}) \valpha(\uu{W}).
\end{equation}
Multiplying Eq. (\ref{eq:smallalphasoln}) by $(4/3)G(\uu{W})$ and integrating, we obtain
\begin{equation}\label{eq:srhoeqn}
\vrho = \frac{2K}{3} \int (s\mathbb{1} - \uu{W})^{-1} \vrho d\uu{W}.
\end{equation}
Note that in three dimensions the linear transformation $\uu{W}\vx$ can be represented as the cross product $\vom\times\vx$. Without loss of generality we may choose a basis that block-diagonalizes $\uu{W}$, corresponding to the choice of $\vom = \om \hat z$ and
\begin{equation*}
\uu{W} = \om \begin{pmatrix} 0&-1&0\\1&0&0\\0&0&0 \end{pmatrix}.
\end{equation*}
Thus,
\begin{equation*}
(s\mathbb{1} - \uu{W})^{-1} = \
\begin{pmatrix} 
\frac{s}{s^2 + \om^2} & \frac{\om}{s^2 + \om^2} & 0 \\
\frac{-\om}{s^2 + \om^2} & \frac{s}{s^2 + \om^2} &0 \\
0&0&\frac{1}{s}
\end{pmatrix}.
\end{equation*}
This can now be inserted into Eq. (\ref{eq:srhoeqn}), and written in a basis independent format as 
\begin{dmath}\label{eq:rhoominteqn}
\vrho = \int G(\vom) \frac{2K}{3}\left(\frac{(\vrho\cdot\haom)\haom}{s} + \frac{\vom\times\vrho}{s^2+\om^2} \\ + \frac{s}{s^2+\om^2} (\vrho-(\vrho\cdot\haom)\haom)\right) d\vom.
\end{dmath}
We choose the distribution $G(\vom)$ to be an \emph{isotropic} distribution (i.e., a distribution that is invariant to orthogonal transformations) which can hence be written as $G(\vom)d\vom=g(\om)U(\haom)d\om d\haom$, where $U(\haom)=1/(4\pi)$ represents the isotropic distribution of rotation directions, and $g(\om)$ is the distribution of the magnitudes of rotation (see Ref. \onlinecite{Chandra2019} for further discussion on the choice of this distribution and its implications). Integrating over the rotation directions $\haom$ in Eq. (\ref{eq:rhoominteqn}) gives us
\begin{equation}
1 = \frac{2K}{3}\left(\frac{1}{3s} + \frac{2s}{3}\int\frac{g(\om)d\om}{s^2 + \om^2}\right),
\end{equation}
which is identical to the result obtained in Ref. \onlinecite{Chandra2019}. As discussed in Ref. \onlinecite{Chandra2019}, the above equation implies that in the limit of small $K$,
\begin{equation}
s = 2K/9,
\end{equation}
indicating that this completely incoherent state loses stability at $K=0$. Thus using Eq. (\ref{eq:alphaeqn}) allows us to perform the stability analysis of a state easily without having to solve the partial differential equation of the dynamics of the distribution of agents in both $\vx$ and $t$ as was necessary in Ref. \onlinecite{Chandra2019}.

\subsection{Other Examples}

Extensions appropriate to various contexts may be studied using Eq. (\ref{eq:finalansatz}). 

For example, each of the agents in the model described above could have a bias towards a particular subspace, such as birds in a flock that have a preference to align parallel to the surface of the Earth. In this case, the common field of such a system is then defined similar to Eq. (\ref{eq:kuramotorho}) as
\begin{equation}
\vrho(\eta,t) = K [(1-c)\vz + c\uu{\Pi}\vz],
\end{equation}
where $\uu{\Pi}$ is the operator that projects onto the preferred subspace (e.g., if $\hat x$, $\hat y$ and $\hat z$ are unit vectors in rectangular coordinates with $\hat z$ being vertical, then $\uu{\Pi}=\hat x \hat x^T + \hat y \hat y^T$ would represent the preference to align to a horizontal surface), and $0\leq c\leq 1$ models the strength of the preference. Writing $\vz$ using Eq. (\ref{eq:kuramotozintegral}), along with Eq. (\ref{eq:alphaeqn}) then represents the reduced equations for this problem.

Another extension to the Kuramoto model that is often studied is the Kuramoto-Sakaguchi model\cite{Sakaguchi1986}. In this model the $\sin(\theta_j-\theta_i)$ coupling term of the Kuramoto model is replaced with $\sin(\theta_j-\theta_i+\delta)$. A possible generalization  of this to higher dimensions, is represented by defining $\vrho$ as $\vrho = K\uu{R}\vz$, where $\uu{R}$ is a given rotation matrix (for $D=2$, $\uu{R}$ is the two-dimensional rotation matrix that rotates vectors by an angle of $\delta$). 

Another $D$-dimensional generalization whose analysis can be facilitated by the ansatz Eq. (\ref{eq:finalansatz}) is the consideration of time delay, $\vrho(t)=K\vz(t-\tau)$, as studied for $D=2$ in Ref.\onlinecite{Lee2009}.

Also, we note that interactions between multiple communities of Kuramoto-like agents has received attention due to a variety of applications (e.g. Refs. \onlinecite{Barreto2008,Montbrio2004,Abrams2008}), as well as the presence of interesting dynamics, such as chimera states\cite{Abrams2008}. For example, for the case of homogenous natural rotations of $\uu{W}_\xi$ within each community $\xi$,
\begin{equation}
\partial_t \valpha_\xi = \frac{1}{2}(1+|\valpha_\xi|^2)\vrho_{\xi} - (\vrho_{\xi}\cdot\valpha_\xi)\valpha_\xi + \uu{W}_\xi\valpha_\xi,
\end{equation}
where the subscript $\xi$ denotes quantities applying to community $\xi$. For a case of generalizing the Kuramoto model, we define the order parameter $\vz_\xi$ for community $\xi$ as the average orientation of that community, and take $\vrho_{\xi}$ to be 
\begin{equation*}
\vrho_{\xi}=\sum_{\xi'} K_{\xi,\xi'} \vz_{\xi'},
\end{equation*}
with $K_{\xi,\xi'}$ representing the coupling between community $\xi$ and $\xi'$. The order parameters $\vz_\xi$ can be written in terms of $\valpha_\xi$ using Eq. (\ref{eq:kuramotozintegral}) by writing the distribution of rotations for the community $\xi$ as $\delta(\uu{W}-\uu{W}_\xi)$. 

The Kuramoto model with the order parameter defined as Eq. (\ref{eq:ordpar}) is the globally-coupled Kuramoto model, wherein each agent is coupled to every other agent. 
In two dimensions, network-based interaction of agents in Kuramoto-like models have been solved for by an application of the ansatz Eq. (\ref{eq:ansatzpoisson}) for a wide range of network topologies, via a mean-field approach\cite{Restrepo2014,Chandra2017}. An analogous analysis will apply for our generalized ansatz, Eq. (\ref{eq:finalansatz}), for network-based interactions of $D$-dimensional Kuramoto-like units.

\section{Conclusions}\label{sec:conclusions}

There are some strong differences between the case $D=2$ and the case of $D>2$ that must be considered in general. In the case of $D=2$, making the additional assumption that $g(\eta)$ is a suitable analytic distribution of the scalar parameter $\eta$ (e.g., a Lorentzian distribution is often employed), allows the integral in Eq. (\ref{eq:kuramotozintegral}) to be performed via a contour integral, and hence requiring the dynamics of $\valpha(\eta)$ according to Eq. (\ref{eq:alphaeqn}) to be calculated for only one or a few particular complex values of $\eta$\cite{Ott2008}. In $D=2$ this implies that many problems of the form Eq. (\ref{eq:Htcomplex}) with heterogeneous $\eta_i$ reduce to a system of a small number of ordinary differential equations in the $N\to\infty$ limit. For our generalization to higher dimension (where $\eta$ is now a vector parameter with at least two components), we are unable to straightforwardly employ contour integration. Thus, while Eq. (\ref{eq:finalansatz}) represents a strong reduction in the dimensionality of the dynamics as compared to the full system in the $N\to\infty$ limit, i.e., Eq. (\ref{eq:Ddimcontinuity}), it is still not a `low-dimensional system' in the sense of Ref.\onlinecite{Ott2008}, since we must still calculate the dynamics of $\valpha(\eta,t)$ as a function of the vector parameter $\eta$ (as opposed to integrating $\eta$ away via, e.g., a Lorentzian assumption for $g(\eta)$).

For the case of homogenous systems, i.e., where $g(\eta)$ is the Dirac-delta function, the dynamics of the full system Eq. (\ref{eq:model}) reduces to the single $D$ dimensional differential equation Eq. (\ref{eq:alphaeqn}). For the particular case of the Kuramoto model generalized to higher dimensions in the manner given in Sec. \ref{sec:genkura}, this exactly reproduces recent results by Lohe\cite{Lohe2018} for the case of a finite number of homogenous agents derived in the context of a generalization of the Watanabe-Strogatz (WS) transform\cite{Watanabe1993,Watanabe1994}. In 2014 Tanaka\cite{Tanaka2014} considered a generalization of the Kuramoto model with higher dimensional \emph{complex} vectors. In this setup (which is different from ours) he derived an extension of the Ott-Antonsen method in the context of a generalized WS transform. In 2008, Pikovsky and Rosenblum\cite{Pikovsky2011} demonstrated that there is a relationship between the WS transform and the Ott-Antonsen ansatz in the case of the original ($D=2$ in our notation) Kuramoto model. It is possible that similar relationships may exist between our generalization of the Ott-Antonsen ansatz Eqs. (\ref{eq:finalansatz}), (\ref{eq:alphaeqn}) and the generalization of the WS transform described in Refs. \onlinecite{Tanaka2014, Lohe2018} --- we leave the study of this relationship to future work.

In conclusion, we have developed a technique to tackle the generalization of several Kuramoto-like systems into higher dimensions. While our analysis has only demonstrated the existence of an invariant manifold to the dynamics of Eq. (\ref{eq:Ddimcontinuity}), from numerical experiments we observe for all examined examples of systems given Eq. (\ref{eq:model}) with a continuous distribution $g(\eta)$ that this manifold is attracting. That is, initial conditions set up not satisfying Eq. (\ref{eq:finalansatz}) appear to be rapidly attracted towards this invariant manifold. While, in the case of $D=2$, it has been shown analytically that, for a broad class of models of the form given by Eq. (\ref{eq:Htcomplex}), this manifold is a global attractor of the dynamics\cite{Ott2009}, proof of attraction for $D>2$ remains an open problem. Given the wide applicability of Eq. (\ref{eq:Htcomplex}) and its rich variety of dynamical phenomena, we expect that the generalization to higher dimensions, Eq. (\ref{eq:model}), may be a useful model system, applicable to diverse situations of interest, while remaining amenable to analysis via the methods developed in this paper.

\begin{acknowledgments}
This work was supported by ONR grant N000141512134 and by AFOSR grant FA9550-15-1-0171,
\end{acknowledgments}

\appendix
\section{Proof of Eq. (\ref{eq:alphaeqn})}\label{apx:alphaeqn}

Inserting the form of $f(\vx,\eta,t)$ from Eq. (\ref{eq:finalansatz}) into Eq. (\ref{eq:expandedcontinuitywithbeta}) we obtain
\begin{dmath}
\left[(D-1)(1-|\valpha|^2)^{D-2}\right]\left\{ (1+|\valpha|^2 - 2\valpha\cdot \vx)(-2 \valpha\cdot \partial_t \valpha) \\- (1-|\valpha|^2)(2\valpha\cdot \partial_t \valpha - 2\vx\cdot \partial_t \valpha) \\ + (1-|\valpha|^2)\left[ 2(\valpha\cdot\vrho) - (\vrho\cdot\vx)(1+|\valpha|^2) - 2\vx \cdot W\valpha\right]\right\} = 0.
\end{dmath}
Remarkably, the explicit $D$ dependence of the differential equation cancels out, and a differential equation involving only terms that are linear and constant in $\vx$ remains. For this equation to be identically zero for each direction $\vx$, the linear and constant terms must independently be zero. From the constant term we obtain
\begin{dmath*}
(1+|\valpha|^2)(-2 \valpha \cdot \partial_t \valpha) - (1-|\valpha|^2)(2 \valpha \cdot \partial_t \valpha) \\ + (1-|\valpha|^2)(2(\valpha\cdot\vrho)) = 0, 
\end{dmath*}
which simplifies to
\begin{equation}\label{eq:valphadtvalpha}
\valpha \cdot\partial_t \valpha = (1/2)(1-|\valpha|^2)(\vrho\cdot\valpha),
\end{equation}
or alternately
\begin{equation}\label{eq:absalpha}
\partial_t |\valpha| = \left(\frac{1-|\valpha|^2}{2|\valpha|}\right)(\vrho\cdot\valpha).
\end{equation}

From the $\vx$ dependent portion we get
\begin{dmath*}
\vx\cdot\left[2\valpha(2\valpha\cdot\partial_t \valpha) + (1-|\valpha|^2)(2\partial_t \valpha) \\- k(1-|\valpha|^2)(1+|\valpha|^2)\vrho - 2(1-|\valpha|^2) W\valpha   \right]=0.
\end{dmath*}
Since $\vx$ is allowed to be in any direction, we can cancel out the $\vx$ and obtain a vector equation that must be satisfied. To further simplify this vector expression, we write $\partial_t \valpha=\partial_t (|\valpha|\halpha) = |\valpha|\partial_t \halpha + \halpha\partial_t |\valpha|$, where $\halpha$ is a unit vector in the direction of $\valpha$. We can then use Eqs. (\ref{eq:valphadtvalpha}) and (\ref{eq:absalpha}) to simplify the expression to obtain
\begin{equation}\label{eq:hatalpha}
\partial_t \halpha = \left(\frac{1+|\valpha|^2}{2|\valpha|} \right)(\vrho - (\vrho \cdot \halpha)\halpha) + W\halpha.
\end{equation}
Equations (\ref{eq:absalpha}) and (\ref{eq:hatalpha}) can then be combined to obtain Eq. (\ref{eq:alphaeqn}).

\end{document}